\documentclass{elsart}
\usepackage{graphicx,amssymb}

\newcommand{\bb}{beta-beams}
\newcommand{\gv}{{\rm g}_V}
\newcommand{\ga}{{\rm g}_A}
\newcommand{\helium}{$^6{\rm He}$}
\newcommand{\neon}{$^{18}{\rm Ne}$}

\begin{document}
\begin{frontmatter}

\title{Electroweak Tests at Beta-beams}

\author[Madison]{A.B. Balantekin},
\ead{baha@physics.wisc.edu}
\author[Madison]{J.H. de Jesus},
\ead{jhjesus@physics.wisc.edu}
\author[Orsay]{C. Volpe}
\ead{volpe@ipno.in2p3.fr}

\address[Madison]{Department of Physics, University of Wisconsin, Madison, WI 53706, USA}
\address[Orsay]{Institut de Physique Nucl\'eaire, F-91406 Orsay cedex, France}

\begin{abstract}
We explore the possibility of measuring the Weinberg angle from (anti)neutrino-electron scattering using low energy \bb, a method that produces single flavour neutrino beams from the beta-decay of boosted radioactive ions.  We study how the sensitivity of a possible measurement depends on the intensity of the ion beam and on a combination of different Lorentz boosts of the ions.
\end{abstract}
\end{frontmatter}

\section{Introduction}
Soon after electroweak theory was introduced, 't Hooft pointed out that low energy neutrino-electron scattering experiments can be used to test the Standard Model~\cite{'tHooft:1971ht}.  This purely leptonic process measures $\sin^2 \theta_W (Q^2 \sim 0)$. The first such measurement was reported in Ref.~\cite{Gurr:1972pk}, yielding $\sin^2 \theta_W = 0.29 \pm 0.05$.  Two subsequent low $Q^2$ experiments were performed, namely the atomic parity violation at $Q^2 \sim 10^{-10}$ GeV$^2$~\cite{Bennett:1999pd} and the M{\o}ller scattering at $Q^2 = 0.026$ GeV$^2$~\cite{Anthony:2005pm}.  These experiments, combined with the measurements of $\sin^2\theta_W$ at the $Z^0$ pole~\cite{Group:2005em}, are consistent with the expected running of the weak mixing angle.  However, recent measurement of the neutral- to charged-current ratio in muon antineutrino-nucleon scattering at the NuTEV experiment disagrees with these results by about 3$\sigma$~\cite{Zeller:2001hh}.  A number of ideas were put forward to explain the so-called NuTEV anomaly, including QCD and nuclear physics effects, extra U(1) gauge bosons, dimension six-operators~\cite{Davidson:2001ji}; universal suppression of Z-neutrino couplings~\cite{Loinaz:2004qc}; higher twist effects arising from nuclear shadowing~\cite{Miller:2002xh}; and sterile neutrino mixing~\cite{Giunti:2002nh}.  However, a complete understanding of the physics behind the NuTEV anomaly is still lacking.  Probing the Weinberg angle through additional experiments with different systematic errors would be very useful.     

The precision tests of the electroweak theory, in principle, can help determine possible ``oblique corrections'' arising from vacuum polarization corrections with the new particles in the loops and suppressed vertex corrections.  Such corrections can be characterized with two parameters, named S and T~\cite{Peskin:1991sw}.  It was recently stressed that a measurement of the electron antineutrino-electron elastic scattering count rates at 1 or 2\% level would restrict the parameter S more closely than measurements of atomic parity violation~\cite{Rosner:2004yt}.

Motivated in part by the NuTEV result, a strategy was presented in Ref.~\cite{Conrad:2004gw} for measuring $\sin^2 \theta_W$ to about one percent at a reactor-based experiment.  In this article we explore an alternative scenario, namely using low energy \bb.  These are pure beams of electron neutrinos or antineutrinos produced through the decay of radioactive ions circulating in a storage ring~\cite{Zucchelli:2002sa,cernweb}.  In our analysis, we investigate the possibility of using a low energy \bb~facility~\cite{Volpe:2003fi} to carry out such a test, through scattering on electrons at $Q^2 \sim 10^{-4}$~GeV$^2$.

We should point out that several other measurements of the Weinberg angle using neutrino-electron scattering already exist.  These use either conventional muon neutrino beams ($Q^2 \sim 0.01$ GeV$^2$)~\cite{Ahrens:1990fp,Vilain:1994qy} or electron neutrinos coming from muon decay at rest~\cite{Auerbach:2001wg}.  These experiments typically determine $\sin^2 \theta_W$ to an accuracy of five to twenty percent. 

In Section 2 we present the calculations and describe the advantages of using \bb. In Section 3, the selection of the events is presented, as well as the results for the $\Delta \chi^2$ fits and for the $1\sigma$ uncertainty on the Weinberg angle, including both statistical and systematic errors.  The sensitivity of these results to the intensity of the ions in the storage ring is also discussed.  Finally, conclusions are drawn in Section 4.

\section{Calculations}
The differential cross section for $\nu_e(\overline{\nu}_e)e^- \rightarrow \nu_e(\overline{\nu}_e)e^-$ in units of $\hbar c=1$ is~\cite{Vogel:1989iv}
\begin{equation}
\!\!\!\!\!\frac{d\sigma}{dT} ~=~ \frac{G_F^2m_e}{2\pi}\!\left[(\gv+\ga)^2\!+(\gv-\ga)^2\left(1-\frac{T}{E_\nu}\right)^2\!\!+(\ga^2-\gv^2)\,\frac{m_eT}{E_\nu^2}\right]~,
\label{difsigma}
\end{equation}
where\footnote{When considering oblique corrections, $\gv$ and $\ga$ include terms that depend on the parameters $S$ and $T$~\cite{Peskin:1991sw}.} $\gv=1/2+2\sin^2\theta_W$, and $\ga=\pm 1/2$ for $\nu_e$ ($\overline{\nu}_e$), $m_e$ is the electron mass, $G_F$ is the Fermi weak coupling constant, $E_\nu$ is the impinging (anti)neutrino energy and $T$ is the electron (positron) recoil energy.  By integrating Eq.~(\ref{difsigma}) over the electron recoil energy and averaging over the neutrino flux one gets the flux-averaged cross section
\begin{equation}
\langle\sigma\rangle ~=~ \frac{G_F^2m_e}{2\pi}\left[\frac{4}{3} \left(\gv^2+\ga^2+\gv\ga\right)\langle E_\nu\rangle-\gv(\gv+\ga)m_e\langle\phi\rangle\right]~, 
\label{avesigma}
\end{equation}
where we defined
\begin{equation}
\langle E_\nu \rangle ~=~ \int dE_\nu \phi(E_\nu) E_\nu~,
\label{aveenergy}
\end{equation}
and
\begin{equation}
\langle \phi  \rangle ~=~ \int dE_\nu \phi(E_\nu)~.
\label{aveflux}
\end{equation}

We consider a scenario where the (anti)neutrinos are produced by low energy \bb.  The (anti)neutrino flux $\phi(E_\nu)$ is then the one associated to the decay of boosted radioactive ions.  The details of the formalism used, including the calculation of the flux, can be found in~\cite{Serreau:2004kx}.  Since the work in~\cite{Serreau:2004kx} has shown that a small storage ring is more appropriate, we consider such a ring, the actual dimensions being determined according to~\cite{chance}.  We assume that a detector of cylindrical shape, having radius $R$ and depth $h$, is located at a distance $d$ from the storage ring. Therefore, the count rate is given by
\begin{equation}
\frac{dN(\gamma)}{dt} ~=~f \tau nh~\langle\sigma\rangle~.
\label{countrate}
\end{equation}
Here $\gamma$ is the Lorentz boost of the ions, $\tau=t_{1/2}/\ln 2$ is  the lifetime of the parent nuclei, $n$ is the number of target particles per unit volume, and $f$ is the number of injected ions per unit time.  Note that the mean number of ions in the storage ring is $\gamma\tau f$.  Combining Eqs.~(\ref{avesigma})~and~(\ref{countrate}), one gets
\begin{equation}
N(\gamma)E_0(\gamma) ~=~ -\gv\left(\gv+\ga\right)m_e+\frac{4}{3}\left(\gv^2+\ga^2+\gv\ga\right)\frac{\langle E_\nu(\gamma)\rangle}{\langle \phi(\gamma)\rangle}~, 
\label{linear1}
\end{equation}
where $E_0(\gamma)$ is a quantity with units of energy defined as
\begin{equation}
E_0(\gamma) ~=~ \left[\Delta tf \tau nh\left(\frac{G_F^2m_e}{2\pi}\right)\langle\phi(\gamma)\rangle\right]^{-1}~,
\label{e0}
\end{equation}
with $\Delta t$ being the duration of the measurement at each $\gamma$.

\begin{figure}[t]
\begin{center}
\includegraphics*[width=6.5cm]{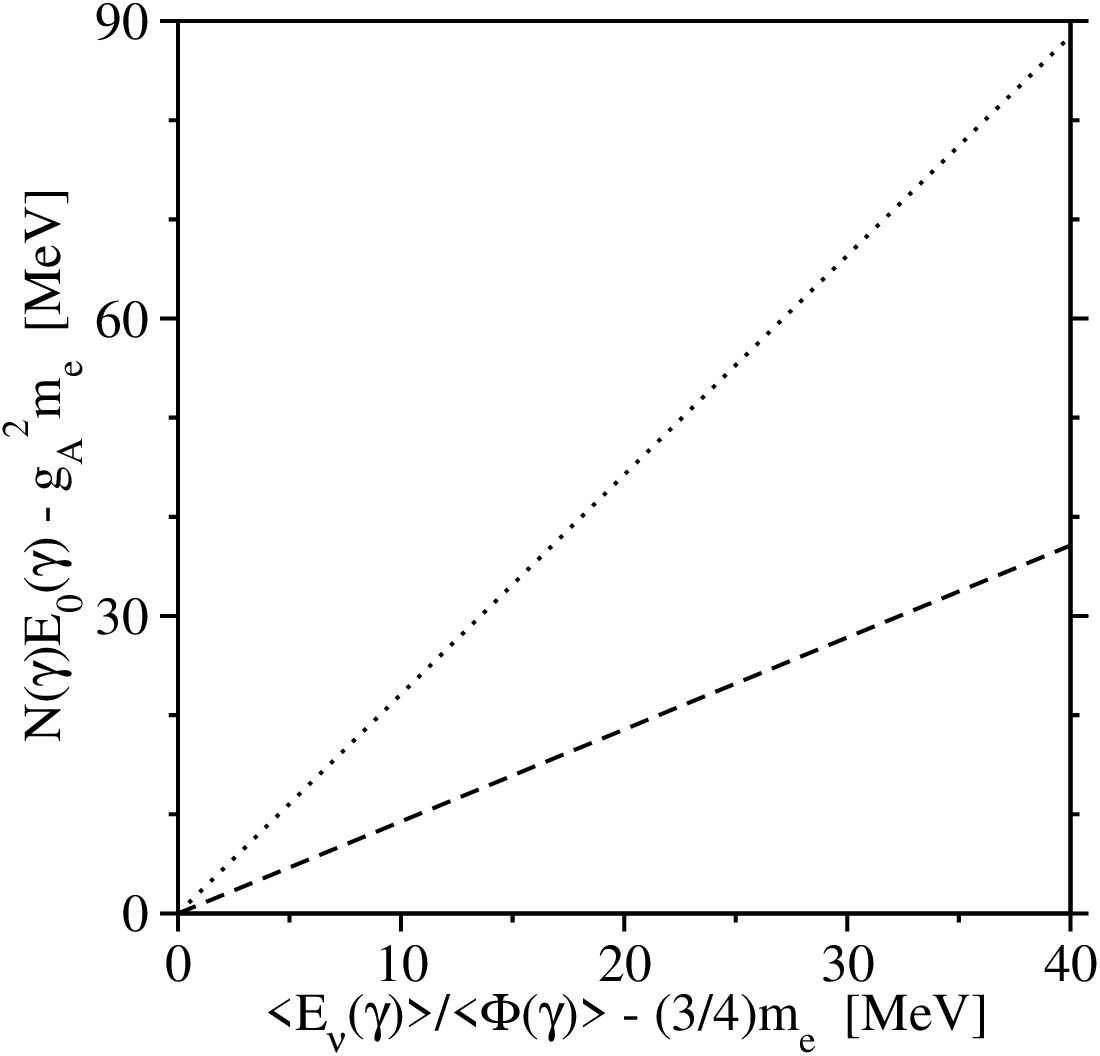}
\end{center}
\caption{Representation of the linear behavior expressed in Eq.~(\ref{linear2}) for electron neutrinos (dotted line) and electron antineutrinos (broken line).  The slope is proportional to $(\gv^2+\ga^2+\gv\ga)$, being different for neutrinos and antineutrinos because of the relative phase in $\ga$.  One measurement of the count number at a particular $\gamma$ for either $(\nu_e,e^-)$ or $(\overline{\nu}_e,e^-)$ scattering is sufficient to determine the Weinberg angle since the y-intercept is zero.}
\label{fig:linear}
\vskip 0.25cm
\end{figure}

Equation~(\ref{linear1}) can be rewritten as
\begin{equation}
N(\gamma)E_0(\gamma)-\ga^2m_e ~=~ \frac{4}{3}\left(\gv^2+\ga^2+\gv\ga\right)\left[\frac{\langle E_\nu(\gamma)\rangle}{\langle \phi(\gamma)\rangle}-\frac{3}{4}m_e\right]~.
\label{linear2}
\end{equation}
If one neglects oblique corrections then $\ga^2=1/2$; in that limit, Eq.~(\ref{linear2}) represents a linear relationship between the number of counts $N(\gamma)$ and the (anti)neutrinos average energy $\langle E(\gamma) \rangle$.  This relationship is depicted in Fig.~\ref{fig:linear}.  The phase difference in $\ga$ is responsible for the difference in the slopes.  Because beta-beam facilities produce pure beams of neutrinos or antineutrinos, one can probe each slope independently and extract information on the Weinberg angle.  In these experiments, the precision in the measurement of the Weinberg angle is determined by the precision to which the slope in Fig.~\ref{fig:linear} is known.  This illustrates the advantage of using beta-beams. First, since the neutrino flux and average energy is well known, the number of counts $N(\gamma)$ is in principle sufficient to extract information on the Weinberg angle.  Second, by changing $\gamma$ one can measure the variation of the cross section with energy, without measuring the electron recoil energies.  Running the experiment at different $\gamma$'s provide a better precision in the determination of the slope in Eq.~(\ref{linear2}), i.e., a better precision in the measurement of the Weinberg angle.

\section{Results and Discussion}
In our calculations we assume a water \v{C}erenkov detector located at 10~m from the storage ring which has 1885~m total length and 678~m straight sections~\cite{chance}. (Note that the present size is intermediate to the ones previously considered \cite{Serreau:2004kx}; the rates given in~\cite{Serreau:2004kx} can be scaled with the analytical formulas given in the same work).  Both electron neutrino and electron antineutrino beams can be produced if \neon~or \helium~are used as emitters.

The expected intensities at production are $8 \times 10^{11}$~ions/s for Neon and $2 \times 10^{13}$~ions/s for Helium, while in the large storage ring envisaged in the original scenario $2 \times 10^{13}$~$^{18}$Ne/s and $4 \times 10^{13}$~$^6$He/s are expected for $\gamma=100$~\cite{Zucchelli:2002sa,cernweb}.  Preliminary calculations show that the intensities in a small storage ring devoted to ions running at $\gamma=7-14$ are $f_{^{18}{\rm Ne}}=0.5 \times 10^{11}$~ions/s and  $f_{^6{\rm He}}=2.7 \times 10^{12}$~ions/s.  However, since the feasibility study of the small storage ring is ongoing, the design we utilize in this work should be considered as preliminary

\begin{table}[t]
\begin{center}
\begin{tabular}{|c|r r|r r|r r|r r c|} \hline
\multicolumn{1}{|c|}{$\langle E(\gamma) \rangle$ (MeV)} & \multicolumn{1}{c}{$N_{\rm tot}$} & \multicolumn{1}{c|}{$\sigma_{\rm tot}$} & \multicolumn{1}{c}{$N_{\rm O}$} & \multicolumn{1}{c|}{$\sigma_{\rm O}$} & \multicolumn{1}{c}{$N_{p}$} & \multicolumn{1}{c|}{$\sigma_{p}$} & \multicolumn{1}{c}{$N_{e}$} & \multicolumn{1}{c}{$\sigma_{e}$} & \multicolumn{1}{c|}{$\sigma_{e}/N_{e}$} \\
 \hline
 22.7 & 115.1 & 10.7 &   6.6 &  2.6 &  44.4 &  6.7 &  64.1 & 12.9 & 20.1 \\
 25.6 & 168.0 & 13.0 &  15.4 &  3.9 &  66.0 &  8.1 &  86.6 & 15.8 & 18.2 \\ 
 28.5 & 236.0 & 15.4 &  31.3 &  5.6 &  92.4 &  9.6 & 112.3 & 19.0 & 16.9 \\
 31.3 & 321.0 & 17.9 &  56.4 &  7.5 & 123.5 & 11.1 & 141.1 & 22.4 & 15.9 \\
 34.2 & 429.1 & 20.7 &  97.1 &  9.9 & 159.0 & 12.6 & 173.0 & 26.2 & 15.1 \\
 37.0 & 557.1 & 23.6 & 150.7 & 12.3 & 198.6 & 14.1 & 207.8 & 30.1 & 14.5 \\
   \hline
\end{tabular}
\end{center}
\caption{Expected number of counts for antineutrino (from the decay of boosted \helium) capture on oxygen, on protons and scattering on electrons and associated statistical errors; for the latter, the relative error (in \%) is also shown.  The average energy of the antineutrinos corresponding to $\gamma=7$ up to $\gamma=12$ is given in the first column, as well as the total expected number of events and the associated statistical error (column two and three).  The duration of the measurement is one year ($3\times 10^7$~s) and the intensity of \helium~at the storage ring is $2.7 \times 10^{12}$~ions/s.}
\label{tab:counts}
\vskip 0.5cm
\end{table}

The antineutrinos impinging on the water detector will interact mainly with the protons, with the oxygen nuclei and with the electrons.  In the case of neutrinos, only the latter two are present.  Since the (anti)neutrino-electron scattering is forward peaked, we apply an angular cut of $\cos\theta > 0.9$.  In the forward direction the $\bar{\nu}-^{16}$O and $\bar{\nu}-p$ events are a small fraction of the total.  Table~\ref{tab:counts} illustrates the expected number of events on electrons, protons and oxygen. We assume that the $\bar{\nu}-p$ events can be identified by doping the water \v{C}erenkov detector with Gadolinium~\cite{Beacom:2003nk}.

\begin{figure}[t]
\begin{center}
\includegraphics*[width=6.5cm]{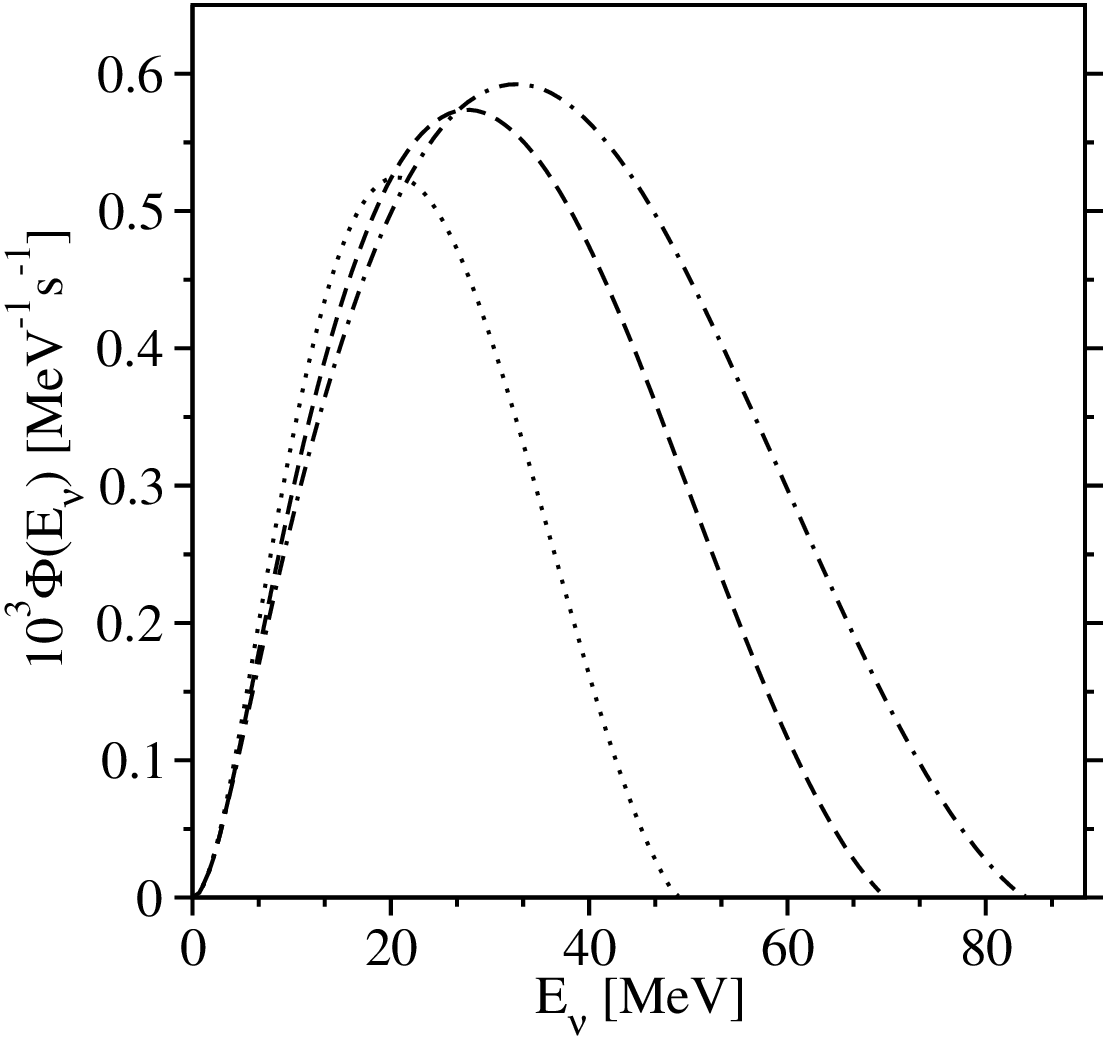}
\end{center}
\caption{Antineutrino flux from the decay of boosted \helium~ions for $\gamma=7$ (dotted line), $\gamma=10$ (broken line) and $\gamma=12$ (dash-dotted line).}
\label{fig:fluxes}
\vskip 0.75cm
\end{figure}

To assess the precision of a possible experiment with \bb, we identified the most critical variables.  If we assume a fixed geometry, Eq.~(\ref{countrate}) tells us that the count rate depends on the boosted ion intensity, $f$, and on the duration of the measurement, $\Delta t$.  On the other hand, the quantity on the left-hand side of Eq.~(\ref{linear2}) is independent of both $f$ and $\Delta t$, since $N(\gamma)$ is proportional to these quantities, while $E_0(\gamma)$ is inversely proportional to them.  This implies that in the diagram of Fig.~\ref{fig:linear}, each experimental data point will be independent of the intensity of the ions and of the duration of the measurement.  The dependence on these quantities, however, contribute to the size of the statistical error, which together with the systematic error determine the precision at which the Weinberg angle can be measured at a low energy beta-beam facility.

The boosted ions that produce neutrinos (\neon) have an intensity fifty times smaller than the ones that produce antineutrinos (\helium)~\cite{chance}.  Even though the scattering cross section is a few times larger for neutrinos than for antineutrinos, it is not sufficient to overcome the big difference in intensities, meaning that for the same duration, an experiment using \helium~will produce considerably better results.

\begin{figure}[t]
\begin{center}
\includegraphics*[width=6.5cm]{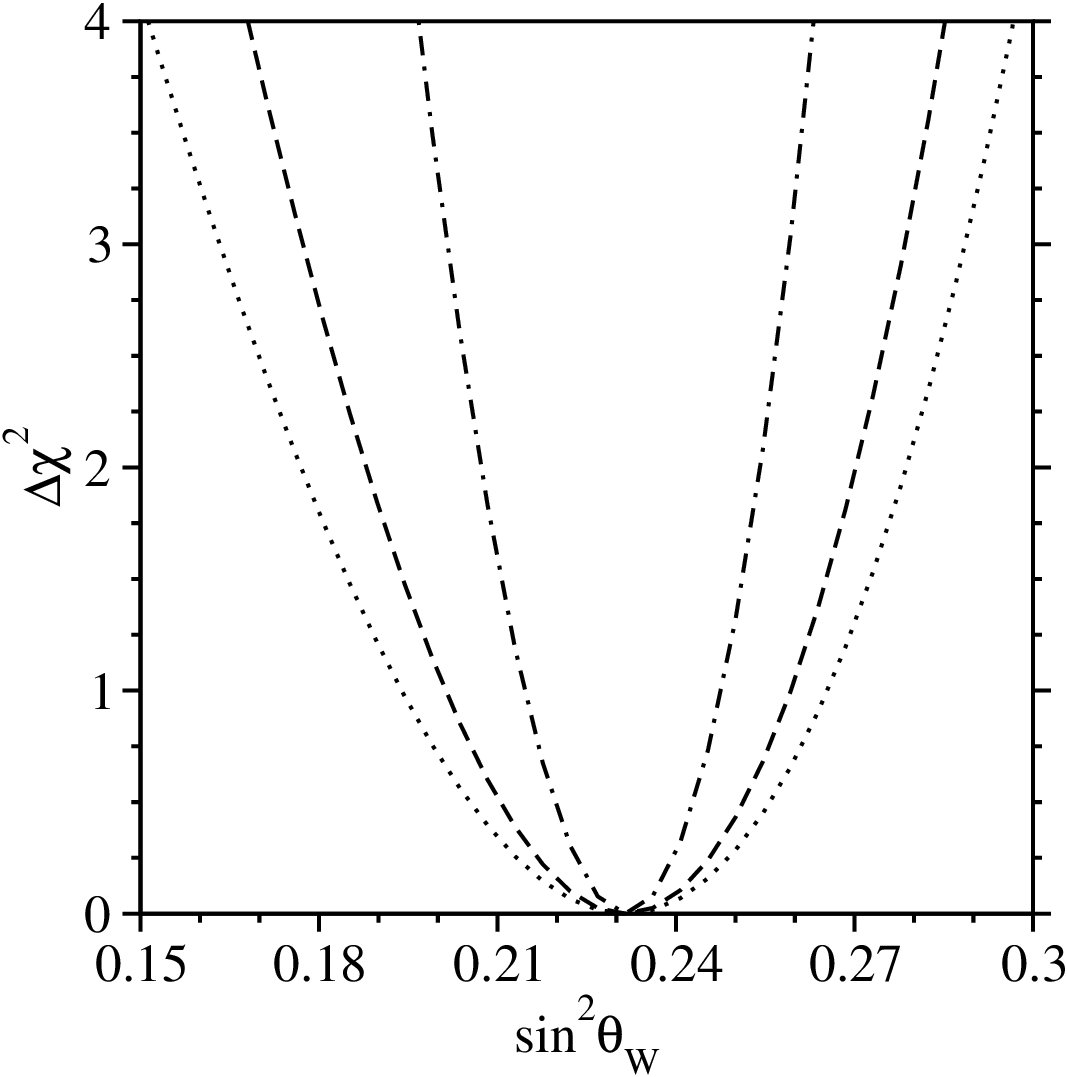}
\end{center}
\caption{$\Delta\chi^2$ with $\gamma=12$ (dotted line), with $\gamma=7,12$ (broken line), and with $\gamma=7,8,9,10,11,12$ (dash-dotted line).  The results were obtained considering a one year ($3 \times 10^7$~s) measurement duration at each $\gamma$, and for a \helium~intensity at the storage ring of $2.7\times 10^{12}$~ions/s~\cite{chance}.  The count number error was considered to be purely statistical.  The 1$\sigma$ ($\Delta \chi^2=1$) relative uncertainty in the Weinberg angle is 15.2\% for $\gamma=12$, 12.3\% for $\gamma=7,12$, and 7.1\% for $\gamma=7,8,9,10,11,12$.}
\vskip 0.75cm
\label{fig:chi2}
\end{figure}

\begin{figure}[t]
\begin{center}
\includegraphics[width=6.5cm]{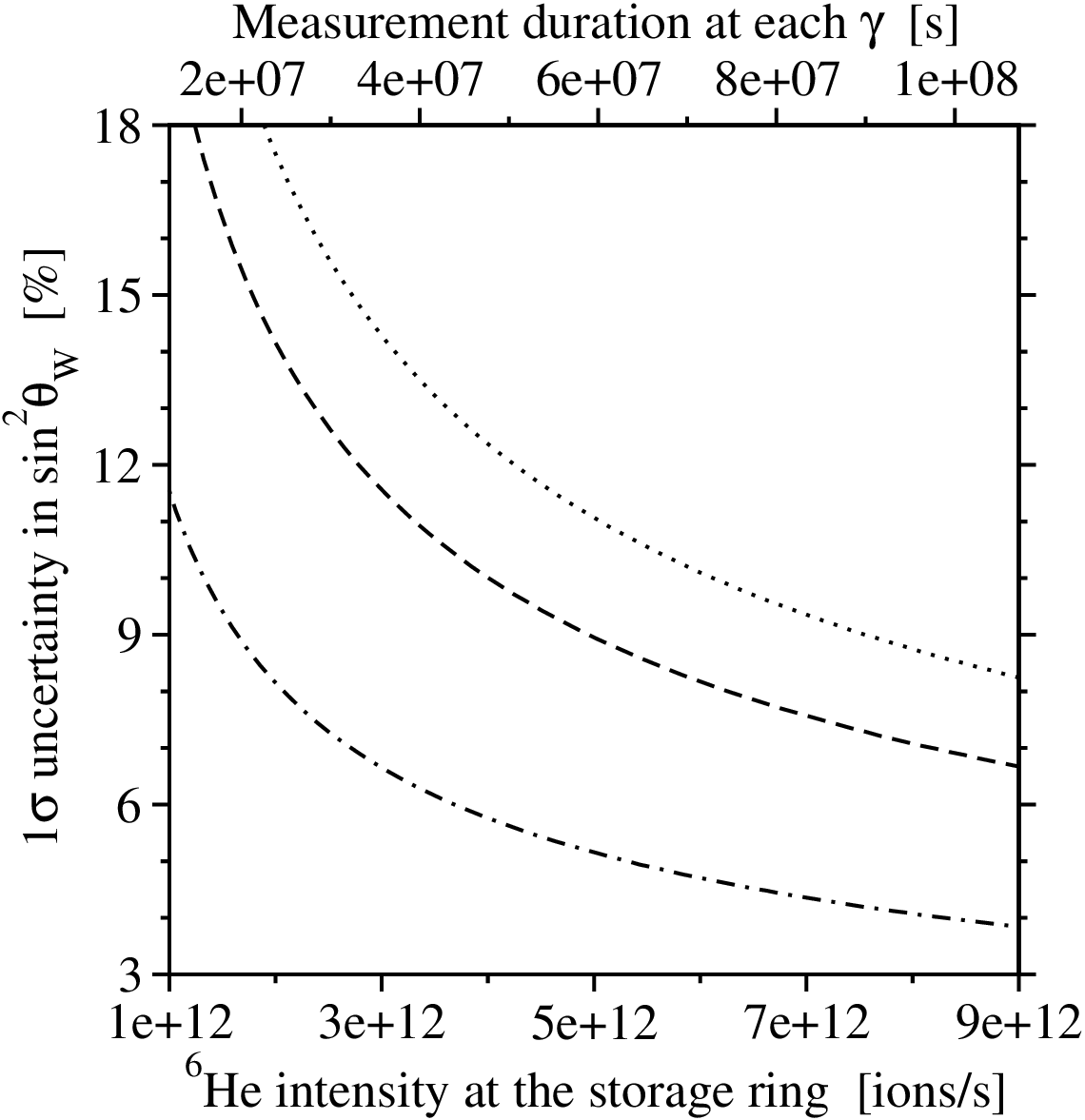}
\end{center}
\caption{One sigma uncertainty in the Weinberg angle as a function of the \helium~intensity at the storage ring (lower x-axis) for a one year ($3\times 10^7$~s) measurement duration at each $\gamma$, and as a function of the measurement duration at each $\gamma$ (upper x-axis) for a \helium~intensity at the storage ring of $2.7\times 10^{12}$~ions/s~\cite{chance}.  Shown are the results for $\gamma=12$ (dotted line), for $\gamma=7,12$ (broken line), and for $\gamma=7,8,9,10,11,12$ (dash-dotted line).  Here, the error on the number of counts is purely statistical.}
\label{fig:sigma}
\vskip 0.75cm
\end{figure}

One advantage of \bb~experiments versus conventional sources or reactors experiments is the ability to produce neutrinos with energies in the 100 MeV range and different average energies.  This means that these experiments are able to span the x-axis of Fig.~\ref{fig:linear} between 20 and 40 MeV, corresponding to $\gamma$ between 7 and 12 (Fig.~\ref{fig:fluxes}).  Running the experiment at a larger number of different $\gamma$'s corresponds to knowing the slope to a better precision, hence measuring the Weinberg angle with better precision.  This is depicted in Fig.~\ref{fig:sigma}, where the 1$\sigma$ error ($\Delta \chi^2=1$) gets smaller as more runs with different $\gamma$ values are added.  For the case of $\gamma=12$, the expected 1$\sigma$ relative uncertainty in the Weinberg angle is 15.2\%, while for $\gamma=7,12$ is 12.3\% and for $\gamma=7,8,9,10,11,12$ is 7.1\%.  These results were obtained assuming a one year ($3 \times 10^7$~s) measurement duration at each $\gamma$, a \helium~intensity at the storage ring of $2.7\times 10^{12}$~ions/s~\cite{chance}, and that the total count error was purely statistical.

The dependence of the uncertainty of the Weinberg angle on the duration of the measurement at each $\gamma$ and on the \helium~intensity at the storage ring is depicted in Fig.~\ref{fig:sigma}; it behaves like $1/\sqrt{f}$ for a fixed duration of the measurement, and like $1/\sqrt{\Delta t}$ for a fixed intensity of the ions.  It is clear that there is much to gain if one increases the intensity of the ions (duration of measurement) by a factor of three compared to the values above, keeping the measurement duration (intensity of the ions) constant.  In that case, the 1$\sigma$ uncertainty in the Weinberg angle drops to 8.8\% for $\gamma=12$, 7.1\% for $\gamma=7,12$ and 4.1\% for $\gamma=7,8,9,10,11,12$.

\begin{figure}[t]
\begin{center}
\begin{minipage}{6.8cm}
\includegraphics[width=6.5cm]{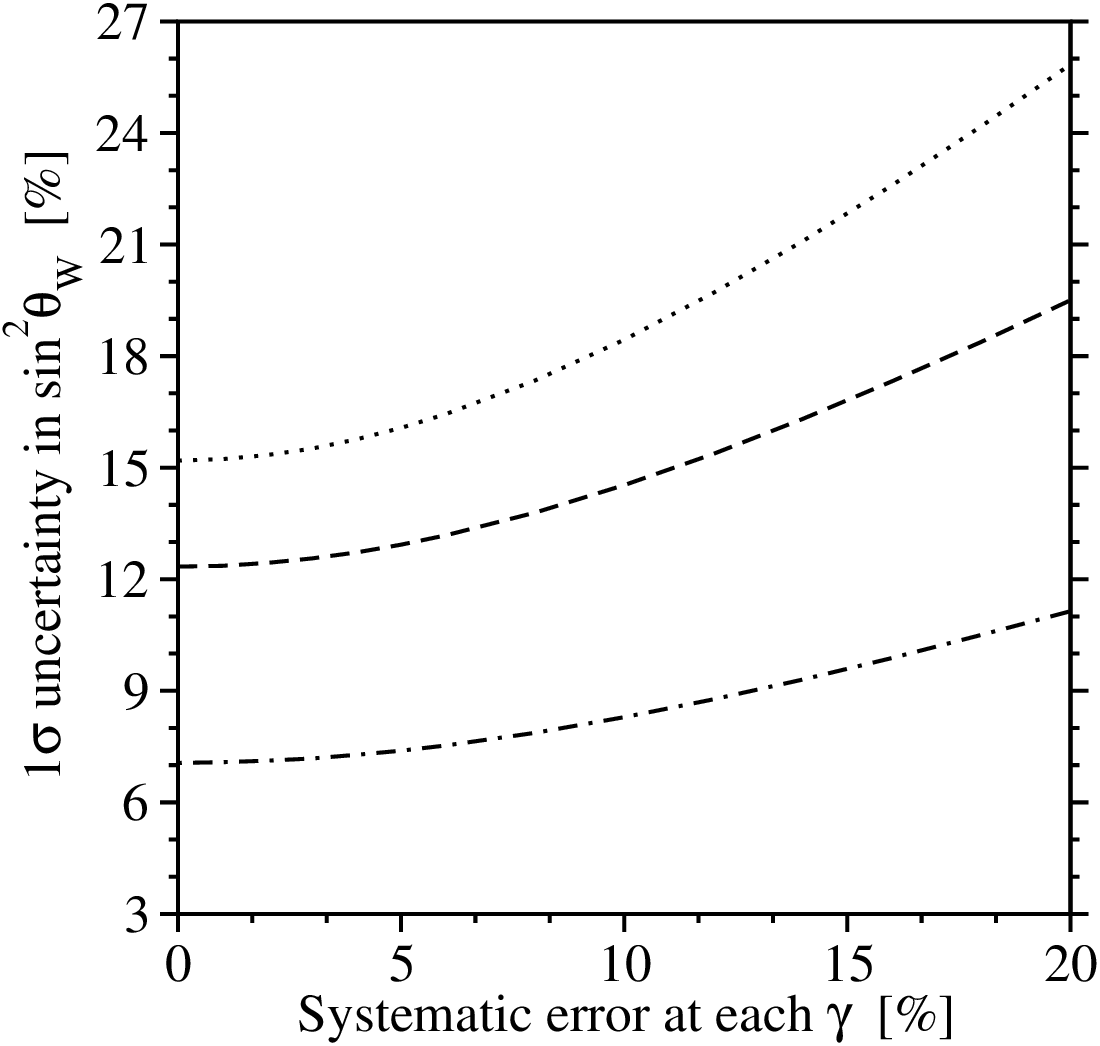}
\end{minipage}
\begin{minipage}{6.8cm}
\includegraphics[width=6.5cm]{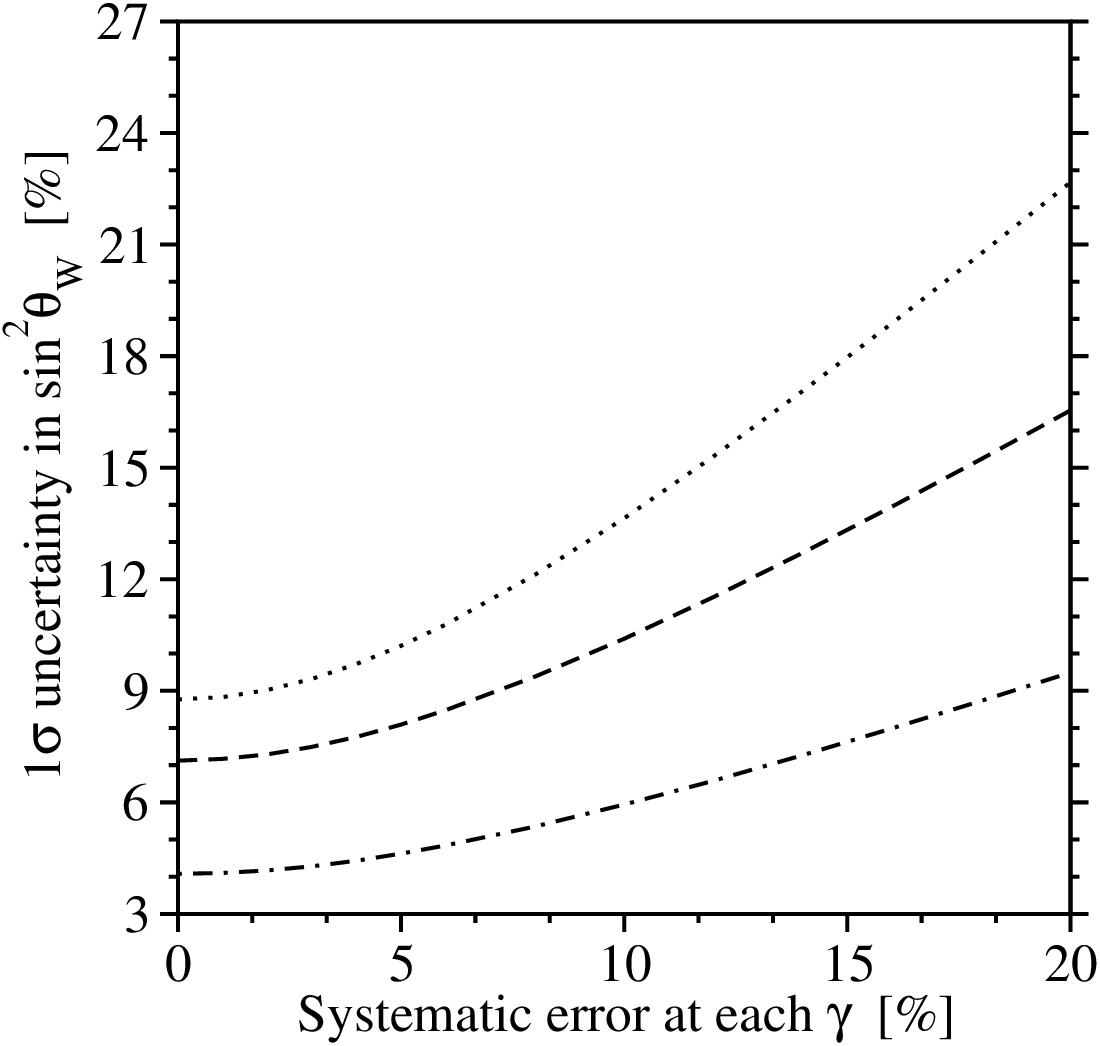}
\end{minipage}
\end{center}
\caption{One sigma uncertainty in the Weinberg angle as a function of the systematic error at each $\gamma$ for $\gamma=12$ (dotted line), for $\gamma=7,12$ (broken line), and for $\gamma=7,8,9,10,11,12$ (dash-dotted line).  The \helium~intensity at the storage ring is $2.7\times 10^{12}$~ions/s~\cite{chance} on the left panel, and $8.1\times 10^{12}$~ions/s on the right panel.  In both cases, the measurement duration at each $\gamma$ is one year ($3\times 10^7$~s).}
\label{fig:syst}
\vskip 0.75cm
\end{figure}

So far we have considered that the error in the number of counts is purely statistical.  Even though one cannot yet know precisely what level of systematic error a beta-beam experiment would produce, it is possible to study its effects on the uncertainty of the Weinberg angle.  In Fig.~\ref{fig:syst}, one can see a considerable dependence of the results on the systematic error, both in the case where the intensity of the ions is the one expected from preliminary studies~\cite{chance} and when it is three times that value.  From this, it is clear that the experiment should aim to minimize the level of systematic errors.

\section{Conclusions}
In this work we have studied the possibility to measure the Weinberg angle at low momentum transfer from (anti)neutrino electron scattering.  The single flavour neutrino fluxes produced at low energy \bb~are very well known. Recent work has pointed to the advantage of using a small storage ring for this kind of applications.  A feasibility study for such facility is now ongoing.  In our analysis, we use the predicted ion intensities from a preliminary study and study how the precision of a possible measurement depends on those intensities as on a combination of different Lorentz boosts for the ions.

We conclude that a measurement of the Weinberg angle at $Q^2\sim 10^{-4}$~GeV$^2$ is possible using low energy \bb.  The level of uncertainty in such a measurement depends strongly on the intensity of the ions at the storage ring, as well as on the duration of the measurement at each $\gamma$ and on the level of systematic errors.  If the intensity of \helium~is a factor of three larger than what the preliminary studies predict, a two years measurement at two different $\gamma$'s would allow to measure the Weinberg angle to a 7\% level, assuming minimal systematic error.  If the systematic error is kept below 10\%, a measurement of the Weinberg angle with a precision of 10\% is within reach at a low energy beta-beam facility.

\section*{Acknowledgments}
We thank M. Benedikt, A. Chanc\'e, M. Lindroos and J. Payet for useful discussion on the feasibility of the storage ring.  We also thank L. Knutson for the insight on data analysis techniques.  The authors acknowledge the CNRS-Etats Units 2005 grant which has been used during the completion of this work.  This work was also supported in part by the U.S. National Science Foundation Grant No. PHY-0244384 at the University of Wisconsin, and in part by the University of Wisconsin Research Committee with funds granted by the Wisconsin Alumni Research Foundation.

\end{document}